\documentclass[aps,preprint,nofootinbib,showpacs]{revtex4}
\usepackage{graphicx}
\usepackage{amssymb}
\usepackage{color}
\def\be{\begin{equation}}
\def\ee{\end{equation}}
\def\bea{\begin{eqnarray}}
\def\eea{\end{eqnarray}}
\def\mz{\mathbb Z}
\def\nn{\nonumber}

\begin{document}

\vspace*{0.2in}
\title{Constructing 5d orbifold grand unified theories\\
from heterotic strings}
\author{Tatsuo Kobayashi,$^a$ Stuart Raby,$^b$ and
Ren-Jie Zhang$^c$\footnote{Also at Institute of Theoretical Physics, The Chinese Academy of Sciences, Beijing 100080, China.}}
\affiliation{$^a$Department of Physics, Kyoto University, Kyoto
606-8502, JAPAN\\
$^b$Department of Physics, The Ohio State University, Columbus, OH
43210 USA\\
$^c$Michigan Center for Theoretical Physics, Randall Laboratory,
University of Michigan, Ann Arbor, MI 48109 USA}

\date{March 5, 2004}

\begin{abstract}
A three-generation Pati-Salam model is constructed by compactifying the heterotic string on a particular
$T^6/{\mathbb Z}_6$ Abelian symmetric orbifold with two discrete Wilson lines. The compactified space is taken to
be the Lie algebra lattice $G_2\oplus SU(3)\oplus SO(4)$.  When one dimension of the $SO(4)$ lattice is large
compared to the string scale, this model reproduces many features of a 5d
$SO(10)$ grand unified
theory compactified on an $S^1/\mz_2$ orbifold.
(Of course, with two large extra dimensions we can obtain
a 6d $SO(10)$ grand unified theory.)
We identify the orbifold parities and other ingredients of the orbifold grand
unified theories in the string model.  Our construction provides a UV completion of orbifold grand unified theories, and gives new
insights into both field theoretical and string theoretical constructions.
\end{abstract}
\preprint{KUNS-1901, MCTP-04-12, OHSTPY-HEP-T-04-003}
\preprint{hep-ph/0403065}

\maketitle

{\bf 1. Motivation}. Particle physics models based on higher-dimensional field theories compactified on orbifolds
have attracted much attention recently \cite{kawamura}. 5d \cite{OGUT} and 6d \cite{6dOGUT} versions of an
$SO(10)$ grand unified theory (GUT) have been studied.  These theories offer novel solutions to some outstanding
problems in conventional 4d GUTs. For example, they allow GUT symmetry breaking without adjoint scalars and
complicated GUT breaking sectors; and they have natural doublet-triplet Higgs splitting, while eliminating
dimension-5 operator contributions to proton decay. However, higher dimensional theories are non-renormalizable
and require an explicit cutoff in order to regularize all the divergences. Moreover, any ultra-violet (UV)
completion of these theories necessarily introduces new physics at the cutoff scale, which will certainly be
relevant for understanding gauge coupling unification, proton decay rates, and family hierarchies.

In order to address these issues, it is essential to obtain a UV completion which is highly motivated in its own
right -- in particular, string theory. Orbifold compactifications \cite{Dixon} of heterotic string theory
\cite{het} have all the necessary ingredients of orbifold GUTs. This motivates us to embed the model in heterotic
string theory. In this Letter, we explicitly construct a three-generation Pati-Salam (PS) model from the
heterotic string compactified on a $T^6/{\mathbb Z}_6$ Abelian symmetric orbifold with two discrete Wilson lines.
(The $T^6/\mz_6$ orbifold under consideration is equivalent to a $T^6/(\mz_2\times\mz_3)$ orbifold.  Note, in
order to reproduce the recent 5d (and 6d) orbifold GUTs, the discrete orbifold point group needs to have a
$\mz_2$ sub-orbifold action.) Our string model is the first three-generation PS model based on non-prime-order
orbifold constructions.\footnote{For a three-generation PS model based on the free fermionic construction, see
Ref.~\cite{antoniadis}.} We reinterpret this model in the orbifold GUT field theory language. Specifically, we
represent the orbifold parities in terms of string theoretical quantities, and identify various
untwisted/twisted-sector states of the string model as bulk/brane states in the orbifold GUT. The main objective
of this Letter is establishing the orbifold GUT--heterotic string connection; details of our model and some
additional three-generation PS models will be presented in a separate publication \cite{KRZ}.

{\bf 2. A 5d orbifold GUT field theory} \cite{OGUT}. The relevant fields under consideration are the gauge field,
taken to be a 5d vector multiplet, ${\cal V}=(V_M,\lambda,\lambda',\sigma)$ (where $V_M$, $\lambda$, $\lambda'$
and $\sigma$ are in the adjoint representations, ${\bf 45}$), and the Higgs field, taken to be a 5d ${\cal N}=2$
hypermultiplet, ${\cal H}=(\phi,\phi^c,\psi,\psi^c)$ (where $\phi$, $\phi^c$, ($\psi$, $\psi^c$) are bosons
(fermions) in the ${\bf 10}+{\bf\overline{10}}$ representation. For $SO(10)$, ${\bf 10}\equiv{\bf\overline{10}}$).
These states are the bulk states in the terminology of 5d theories.
When compactified on a smooth manifold such as the circle, $S^1$, with radius $R$, the above 5d GUT model results
in a 4d $SO(10)$ model with (extended) ${\cal N}=2$ supersymmetry. For every 4d state, there is a tower of ${\cal
N}=2$ Kaluza-Klein (KK) excitations in the same group representation with mass $m/R$ (where
the non-negative integers $m$ label the KK levels). It is often more convenient to write the ${\cal N}=2$ multiplets
in terms of ${\cal N}=1$ multiplets. In the $SO(10)$ model, the 4d massless states are a vector multiplet,
$V=(A_\mu,\lambda)$, a chiral multiplet, $\Sigma=((\sigma+iA_5)/\sqrt{2},\lambda')$, both in the adjoint
representation, and a pair of chiral multiplets, $H=(\phi,\psi)$ and $H^c=(\phi^c,\psi^c)$, in complex conjugate
representations.

The 4d effective theory is quite different, however, if the compactified space is an orbifold
instead of a smooth manifold. Then not only can the extended supersymmetry be broken (partially or completely)
but the GUT gauge group can also be reduced by non-trivial embeddings of the orbifold action into the gauge
degrees of freedom.

Consider the $SO(10)$ example and take the extra dimension to be
an orbi-circle $S^1/\mz^2$. The space group of this orbifold is
generated by two actions, a space reversal, ${\cal P}:\,
y\rightarrow -y$, and a lattice translation, ${\cal
T}:\,y\rightarrow y+2\pi R$. The translation can be replaced by an
equivalent ${\mathbb Z}_2$ action, ${\cal P'}={\cal PT}$. The
fundamental region of $S^1/\mz_2$ is the interval $[0,\,\pi R]$,
where the two ends, $y=0$ and $y=\pi R$ are the fixed points of
${\cal P}$ and ${\cal P}'$.   The orbifold actions ${\cal P}$ and
${\cal P}'$ can be realized on a generic 5d field as orbifold
parities, $P,\,P'=\pm$. Let us assign the following parities to the
fields in the $SO(10)$ model (where we have written the fields in
representations of the PS group, $SU(4)\times SU(2)_L\times
SU(2)_R$), 
\be
\begin{array}{l|ccc|cl|cc}
 {\rm States} & P & P' & & & {\rm States} & P & P'\\
\hline
V({\bf 15,1,1}) & + & + & & &\Sigma({\bf 15,1,1}) & - & -\\
V({\bf 1,3,1})  & + & + & & &\Sigma({\bf 1,3,1}) & - & -\\
V({\bf 1,1,3})  & + & + & & &\Sigma({\bf 1,1,3}) & - & -\\
V({\bf 6,2,2})  & + & - & & &\Sigma({\bf 6,2,2}) & - & +\\
H({\bf 6,1,1})  & + & - & & &H^c({\bf 6,1,1}) & - & + \\
H({\bf 1,2,2})  & + & + & & &H^c({\bf 1,2,2}) & - & -
\end{array}\,.
\ee
The first orbifold parity, $P$, preserves the $SO(10)$ symmetry; its fixed point at $y=0$ is the ``$SO(10)$ brane".
The second projection, $P'$, breaks the $SO(10)$ gauge symmetry to the PS gauge
group; its fixed point at $y=\pi R$ is the ``PS brane".

Masses of KK excitations of these fields depend on their parities,
\be
M_{KK}=\left\{\begin{array}{ll}
m/R & \quad {\rm for}\,\, P=P'=+\,,\\
(2m+1)/2R & \quad {\rm for}\,\, P=+,P'=-\,\,{\rm and}\,\,P=-,P'=+\,, \\
(m+1)/R & \quad {\rm for}\,\,P=P'=-\,.\end{array}\right.\label{KKmass}
\ee
The 4d effective theory includes only zero modes with $P=P'=+$. 
They are the PS gauge fields and the $H({\bf 1,2,2})$
chiral multiplet (which is the minimal supersymmetric standard
model (MSSM) Higgs doublet). Zero modes of the $H({\bf 6,1,1})$
and $H^c({\bf 6,1,1})$ states (which are the MSSM color triplet
Higgses) are absent; this solves the doublet-triplet splitting
problem that plagues conventional 4d GUT theories.

{\bf 3. Heterotic string compactified on $T^6/\mz_6$}.
Let us denote the $\mz_6$ action on the three complex compactified coordinates by $Z^i\rightarrow e^{2\pi i {\bf
r}_i\cdot{\bf v}_6}Z^i$, $i=1,2,3$, where ${\bf v}_6=\frac{1}{6}(1,2,-3)$ is the twist vector, and ${\bf
r}_1=(1,0,0,0)$, ${\bf r}_2=(0,1,0,0)$, ${\bf r}_3=(0,0,1,0)$.\footnote{Together with ${\bf r}_4=(0,0,0,1)$, they
form the set of positive weights of the ${\bf 8}_v$ representation of the $SO(8)$, the little group in 10d.
$\pm{\bf r}_4$ represent the two uncompactified dimensions in the light-cone gauge. Their space-time fermionic
partners have weights ${\bf r}=(\pm\frac{1}{2},\pm\frac{1}{2},\pm\frac{1}{2},\pm\frac{1}{2})$ with even numbers
of positive signs; they are in the ${\bf 8}_s$ representation of $SO(8)$. In this notation, the fourth component
of ${\bf v}_6$ is zero.\label{fn1}} For simplicity and definiteness, we also take the compactified space to be a
factorizable Lie algebra lattice $G_2\oplus SU(3)\oplus SO(4)$.

The $\mz_6$ orbifold is equivalent to a $\mz_2\times\mz_3$
orbifold, where the two twist vectors are ${\bf v}_2=3{\bf
v}_6=\frac{1}{2}(1,0,-1)$ and ${\bf v}_3=2{\bf
v}_6=\frac{1}{3}(1,-1,0)$. The $\mz_2$ and $\mz_3$ sub-orbifold
twists have the $SU(3)$ and $SO(4)$ planes as their fixed torii.
In Abelian symmetric orbifolds, gauge embeddings of the point
group elements and lattice translations are realized by shifts of
the momentum vectors, ${\bf P}$, in the $E_8\times E_8$ root
lattice\footnote{The $E_8$ root lattice is given by the set of
states ${\bf P} = \{n_1, n_2, \cdots, n_8 \}, \ \{n_1
+\frac{1}{2}, n_2 +\frac{1}{2}, \cdots, n_8 +\frac{1}{2} \}$
satisfying $n_i \in \mz, \ \sum_{i = 1}^8 n_i = 2
\mz$.}~\cite{IMNQ}, {\it i.e.}, ${\bf P}\rightarrow {\bf P}+k{\bf
V}+l{\bf W}$, where $k, l$ are some integers, and ${\bf V}$ and
${\bf W}$ are known as the gauge twists and Wilson lines
\cite{wl}. These embeddings are subject to modular invariance
requirements \cite{Dixon,vafa}. The Wilson lines are also required
to be consistent with the action of the point group. In the
$\mz_6$ model, there are at most three consistent Wilson lines
\cite{kobayashi}, one of degree 3 (${\bf W}_3$), along the $SU(3)$
lattice, and two of degree 2 (${\bf W}_2,\,{\bf W}_2'$), along the
$SO(4)$ lattice.

\begin{figure*}
\scalebox{0.85}{\includegraphics{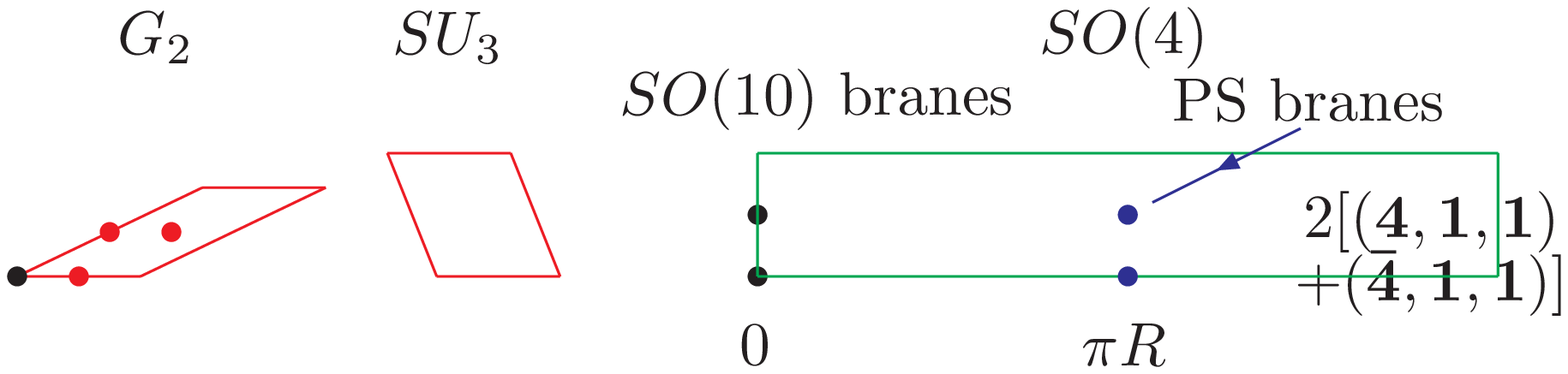}} \caption{$G_2 \oplus
SU(3) \oplus SO(4)$ lattice with $\mz_2$ fixed points. The $T_{3}$
twisted sector states sit at these fixed points.  The fixed point
at the origin and the symmetric linear combination of the red
(grey) fixed points in the $G_2$ torus have $\gamma =1$.
\label{fig:z2} }
\end{figure*}

\begin{figure*}
\scalebox{0.85}{
\includegraphics{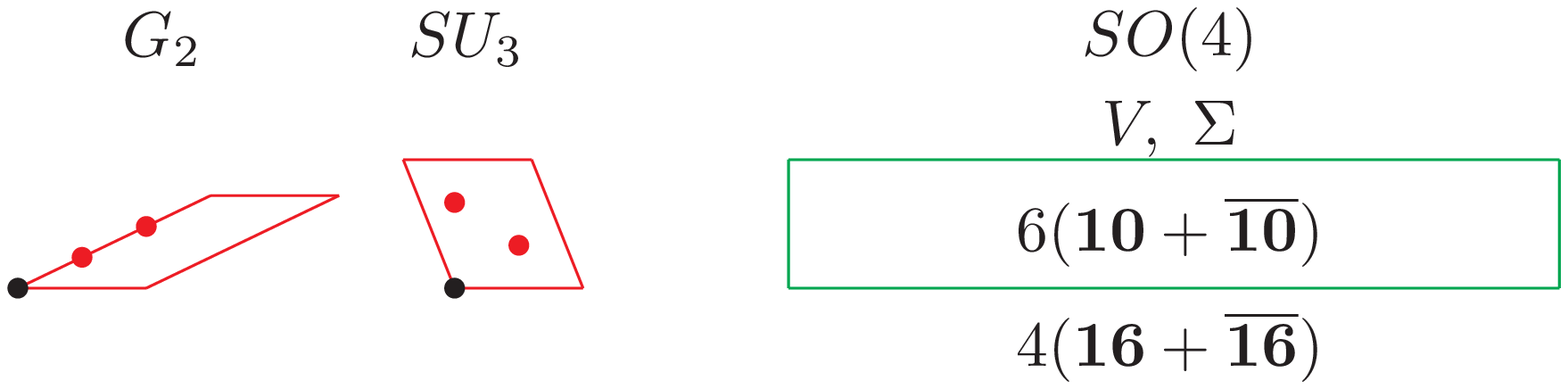}}
\caption{\label{fig:z3} $G_2 \oplus SU(3) \oplus SO(4)$ lattice with $\mz_3$ fixed points.  The fixed point at
the origin and the symmetric linear combination of the red (grey) fixed points in the $G_2$ torus have $\gamma
=1$.  The fields $V, \ \Sigma, \ $ and $1\times({\bf 16} + {\bf\overline{16}})$ are bulk states from the
untwisted sector. On the other hand, $6\times({\bf 10} + {\bf\overline{10}})$ and $3\times({\bf 16} +
{\bf\overline{16}})$ are ``bulk" states located on the $T_{2}/T_{4}$ twisted sector fixed points.}
\end{figure*}

\begin{figure*}
\scalebox{0.85}{
\includegraphics{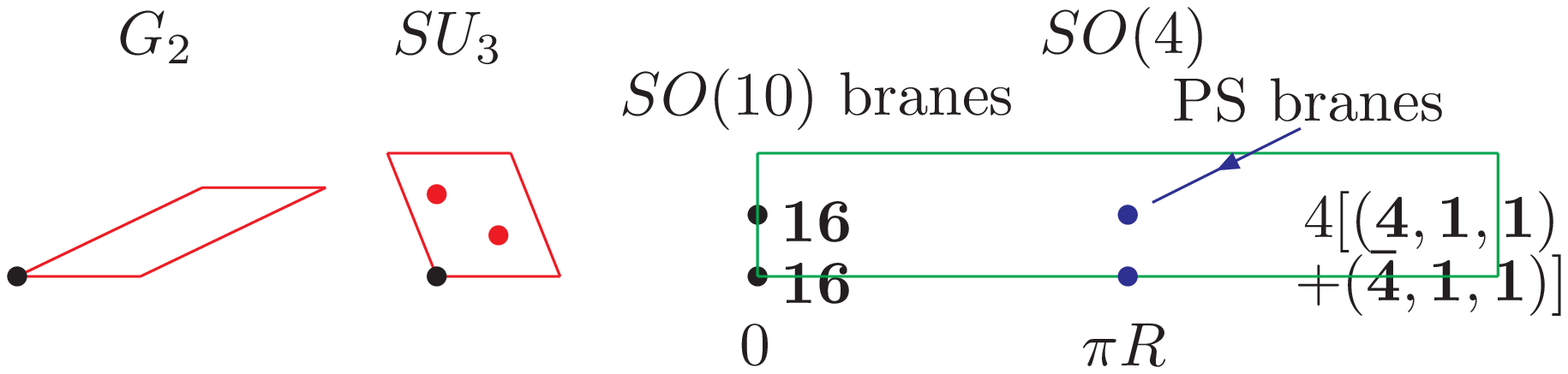}}
\caption{\label{fig:z6} $G_2 \oplus SU(3) \oplus SO(4)$ lattice with $\mz_6$ fixed points. The $T_{1}$ twisted
sector states sit at these fixed points.}
\end{figure*}

The $\mz_6$ model has three untwisted sectors ($U_i,\,i=1,2,3$)
and five twisted sectors ($T_i,\,i=1,2,\cdots,5$). (The $T_k$ and
$T_{6-k}$ sectors are CPT conjugates of each other.) The twisted
sectors split further into sub-sectors when discrete Wilson lines
are present. In the $SU(3)$ and $SO(4)$ directions, we can label
these sub-sectors by their winding numbers, $n_3=0,1,2$ and
$n_2,\,n'_2=0,1$, respectively. In the $G_2$ direction, where both
the $\mz_2$ and $\mz_3$ sub-orbifold twists act, the situation is
more complicated.  There are four $\mz_2$ fixed points in the
$G_2$ plane. Not all of them are invariant under the $\mz_3$
twist, in fact three of them are transformed into each other. Thus
for the $T_3$ twisted-sector states one needs to find linear
combinations of these fixed-point states such that they have
definite eigenvalues, $\gamma=1$ (with multiplicity 2),
$e^{i2\pi/3}$, or $e^{i4\pi/3}$, under the orbifold twist
\cite{DFMS,kobayashi} (see Fig.~\ref{fig:z2}). Similarly, for the
$T_{2,4}$ twisted-sector states, $\gamma=1$ (with multiplicity 2)
and $-1$ (the fixed points of the $T_{2,4}$ twisted sectors in the
$G_2$ torus are shown in Fig.~\ref{fig:z3}). The $T_{1}$
twisted-sector states have only one fixed point in the $G_2$
plane, thus $\gamma=1$ (see Fig.~\ref{fig:z6}). The eigenvalues
$\gamma$ provide another piece of information to differentiate
twisted sub-sectors.


Massless states in 4d string models
consist of those momentum vectors ${\bf P}$ and ${\bf r}$ ($\bf r$ are in the $SO(8)$ weight lattice)
which satisfy the following mass-shell equations
\cite{Dixon,IMNQ}, 
\bea 
&&\frac{\alpha'}{2}m_{R}^2=N^k_R+\frac{1}{2}\left|{\bf r}+k{\bf v}\right|^2+a^k_R=0\,,\label{masscond1}\\
&&\frac{\alpha'}{2}m_L^2=N_L^k+\frac{1}{2}\left|{\bf P}+k{\bf
X}\right|^2+a_L^k=0\,, \label{masscond2} 
\eea 
where $\alpha'$ is the Regge slope, $N^k_R$ and $N^k_L$ are
(fractional) numbers of the right- and left-moving (bosonic)
oscillators, ${\bf X}={\bf V}+n_3{\bf W}_3+n_2{\bf W}_2+n_2'{\bf
W}_2'$, and $a^k_R$, $a_L^k$ are the normal ordering constants, 
\bea 
a^k_R
&=&-\frac{1}{2}+\frac{1}{2}\sum_{i=1}^3|{\widehat{kv_i}}|\left(1-|{\widehat{kv_i}}|\right)\,,\nn\\
a_L^k
&=&-1+\frac{1}{2}\sum_{i=1}^3|{\widehat{kv_i}}|\left(1-|{\widehat{kv_i}}|\right)\,,
\eea 
with $\widehat{kv_i}={\rm mod}(kv_i, 1)$.

These states are subject to a generalized Gliozzi-Scherk-Olive (GSO) projection
${\cal P}=\frac{1}{6}\sum_{\ell=0}^{5}\Delta^\ell$~\cite{IMNQ}. For the simple case of the $k$-th twisted sector ($k=0$ for
the untwisted sectors) with no Wilson lines ($n_3 = n_2 = n^\prime_2 = 0$) we have
\be
\Delta= \gamma\phi  \exp\left\{i\pi \biggl[(2{\bf P}+k{\bf X}) \cdot{\bf X} -(2{\bf r}+k{\bf v})\cdot {\bf
v}\biggr]\right\},\label{GSO}
\ee
where $\phi$ are phases from bosonic oscillators.   However, in
the $\mz_6$ model, the GSO projector must be modified for the
untwisted-sector and $T_{2,4}$, $T_3$ twisted-sector states in the
presence of Wilson lines \cite{KRZ}. The Wilson lines split each
twisted sector into sub-sectors and there must be additional
projections with respect to these sub-sectors. This modification
in the projector gives the following projection conditions,
\be
{\bf P}\cdot{\bf V}-{\bf r}_i\cdot{\bf
v}=\mz\,\,\,\,(i=1,2,3),\quad {\bf P}\cdot{\bf W}_3,\,\,{\bf
P}\cdot{\bf W}_2,\,\,{\bf P}\cdot{\bf W}_2'=\mz, \label{eq1}
\ee
for the untwisted-sector states, and
\be
T_{2,4}:\,{\bf P}\cdot{\bf W}_2,\,\,{\bf P}\cdot{\bf
W}_2'={\mathbb Z}\,,\qquad T_3:\,{\bf P}\cdot{\bf W}_3={\mathbb
Z}\,,\label{eq2}
\ee
for the $T_{2,3,4}$ sector states (since twists of these sectors
have fixed torii). There is no additional condition
for the $T_1$ sector states.

{\bf 4. An orbifold GUT -- heterotic string dictionary}. We first implement the $\mz_3$ sub-orbifold twist, which
acts only on the $G_2$ and $SU(3)$ lattices. The resulting model is a 6d gauge theory with ${\cal N}=2$
hypermultiplet matter, from the untwisted and $T_{2,4}$ twisted sectors. This 6d theory is our starting point to
reproduce the orbifold GUT models.  The next step is to implement the $\mz_2$ sub-orbifold twist. The geometry of
the extra dimensions closely resembles that of the 6d orbifold GUTs. The $SO(4)$ lattice has four $\mz_2$ fixed
points at $0$, $\pi R$, $\pi R^\prime$ 
and $\pi(R + R^\prime)$, where $R$ and $R'$ are the two axes
of the lattice (see Figs.~\ref{fig:z2} and \ref{fig:z6}). When one varies the modulus parameter of the $SO(4)$
lattice such that the length of one axis ($R$) is much larger than the other ($R'$) and the string length scale
($\ell_s$), the lattice effectively becomes the $S^1/\mz_2$ orbi-circle in the 5d orbifold GUT, and the two fixed
points at $0$ and $\pi R$ have degree-2 degeneracies. Furthermore, one may identify the states in the
intermediate $\mz_3$ model, {\it i.e.} those of the untwisted and $T_{2,4}$ twisted sectors, as bulk states in
the orbifold GUTs.

Space-time supersymmetry and GUT breaking in string models work exactly as in the orbifold GUT models.  First consider
supersymmetry breaking. In the field theory, there are two gravitini in 4d, coming from the 5d (or 6d) gravitino.
Only one linear combination is consistent with the space reversal, $y\rightarrow -y$; this breaks the ${\cal
N}=2$ supersymmetry to that of ${\cal N}=1$. In string theory, the space-time supersymmetry currents are
represented by those half-integral $SO(8)$ momenta (see footnote \ref{fn1}). The $\mz_3$ and $\mz_2$
projections remove all but two of them, ${\bf r}=\pm\frac{1}{2}(1,1,1,1)$; this gives ${\cal N}=1$ supersymmetry
in 4d.

Now consider GUT symmetry breaking. As usual, the $\mz_2$ orbifold twist and the translational symmetry of the
$SO(4)$ lattice are realized in the gauge degrees of freedom by degree-2 gauge twists and Wilson lines
respectively. To mimic the 5d orbifold GUT example, we impose only one degree-2 Wilson line, ${\bf W}_2$, along
the long direction of the $SO(4)$ lattice, ${\bf R}$.\footnote{Wilson lines can be used to reduce the number of
chiral families. In all our models, we find it is sufficient to get three-generation models with two Wilson
lines, one of degree 2 and one of degree 3. Note, however, that with two Wilson lines in the $SO(4)$ torus we can
break $SO(10)$ directly to $SU(3) \times SU(2) \times U(1)_Y \times U(1)_X$ (see for example,
Ref.~\cite{6dOGUT}).} The gauge embeddings generally break the 5d/6d (bulk) gauge group further down to its
subgroups, and the symmetry breaking works exactly as in the orbifold GUT models. This can clearly be seen from
the following string theoretical realizations of the orbifold parities
\be P=p\,e^{2\pi i\,[{\bf P}\cdot{\bf V}_2-{\bf r}\cdot{\bf
v}_2]}\,,\quad P'=p\,e^{2\pi i\,[{\bf P}\cdot({\bf V}_2+{\bf
W}_2)-{\bf r}\cdot{\bf v}_2]}\,,\label{PP'}
\ee
where ${\bf V}_2=3{\bf V}_6$, and $p=\gamma\phi$ can be identified
with intrinsic parities in the field theory language.\footnote{For
gauge and untwisted-sector states, $p$ are trivial. For
non-oscillator states in the $T_{2,4}$ twisted sectors, $p=\gamma$
are the eigenvalues of the $G_2$-plane fixed points under the
${\mathbb Z}_2$ twist. Note that $p=+$ and $-$ states have
multiplicities $2$ and $1$ respectively since the corresponding
numbers of fixed points in the $G_2$ plane are $2$ and $1$.} Since
$2({\bf P}\cdot{\bf V}_2-{\bf r}\cdot{\bf v}_2),\,2{\bf
P}\cdot{\bf W}_2=\mz$, by properties of the $E_8\times E_8$ and
$SO(8)$ lattices, thus $P^2=P'^2=1$, and Eq.~(\ref{PP'}) provides
a representation of the orbifold parities. From the string theory
point of view, $P=P'=+$ are nothing but the projection conditions,
$\Delta=1$, for the untwisted and $T_{2,4}$ twisted-sector states
(see Eqs.~(\ref{GSO}), (\ref{eq1}) and (\ref{eq2})).

To reaffirm this identification, we compare the masses of KK
excitations derived from string theory with that of orbifold GUTs.
The coordinates of the $SO(4)$ lattice are untwisted under the
$\mz_3$ action, so their mode expansions are the same as that of
toroidal coordinates. Concentrating on the ${\bf R}$ direction,
the bosonic coordinate is
$X_{L,R}=x^{}_{L,R}+p^{}_{L,R}(\tau\pm\sigma)+{\rm oscillator\,
terms}$, with $p^{}_L$, $p^{}_R$ given by
\be
p^{}_L=\frac{m}{2R}+\left(1-\frac{1}{4}|{\bf W}_2|^2\right)
\frac{n_2R}{\ell_s^2}+\frac{{\bf P}\cdot{\bf W}_2}{2R} \,,\qquad
p^{}_R=p^{}_L-\frac{2n_2R}{\ell_s^2}\,,\label{PLR}
\ee
where $m$ ($n_2$) are KK levels (winding numbers). The ${\mathbb Z}_2$ action maps $m$ to $-m$, $n_2$ to $-n_2$
and ${\bf W}_2$ to $-{\bf W}_2$, so physical states must contain linear combinations,
$|m,n_2\rangle\pm|-m,-n_2\rangle$; the eigenvalues $\pm 1$ correspond to the first ${\mathbb Z}_2$ parity, $P$,
of orbifold GUT models. The second orbifold parity, $P'$, induces a non-trivial degree-2 Wilson line; it shifts
the KK level by $m\rightarrow m+{\bf P}\cdot{\bf W}_2$. Since $2{\bf W}_2$ is a vector of the (integral)
$E_8\times E_8$ lattice, the shift must be an integer or half-integer. When $R\gg R'\sim\ell_s$, the winding
modes and the KK modes in the smaller dimension of $SO(4)$ decouple.  Eq.~(\ref{PLR}) then gives four types of KK
excitations, reproducing the field theoretical mass formula in Eq.~(\ref{KKmass}).

{\bf 5. A three-generation PS model}. To illustrate the above
points, we consider an explicit three-generation PS model in the
$\mz_6$ orbifold, with the following gauge twist and Wilson lines,
\bea
{\bf V}_6&=&\frac{1}{6}\left(222000 00\right)\left(1100
0000\right)\,,\label{gt1} \\
{\bf W}_3&=&\frac{1}{3}\left(2  1  -1  0 0
 0  0  0\right)\left(
 0  2  1  1  0  0  0  0\right)\,,\\
{\bf W}_2&=&\frac{1}{2}\left( 1  0  00
0 1  1  1\right) \left(
0  0 0  0  0  0 0  0\right)\,. \label{wla2}
\eea
The unbroken gauge groups in 4d are $SU(4)\times SU(2)_L\times SU(2)_R\times SO(10)'\times SU(2)'\times U(1)^5$
(one of the Abelian groups is anomalous), and the untwisted- and twisted-sector matter states furnish the
following irreducible representations of the PS gauge group (modulo singlets),
\bea
U_1: && ({\bf 4},{\bf 2}, {\bf 1})
\,,\nonumber\\
T_1: && 2({\bf 4}, {\bf 2}, {\bf 1})+2(\bar{\bf 4},{\bf 1},{\bf
2}) +4({\bf 4},{\bf 1},{\bf 1})+4(\bar{\bf 4},{\bf 1},{\bf
1})+8({\bf 1,2,1})
+6({\bf 1,1,2})\nn\\
&&+2({\bf 1,2,1;1,2})+2({\bf 1,1,2;1,2})\,,\nn\\
T_2: && 2(\bar{\bf 4},{\bf 1},{\bf 2})+({\bf 6},{\bf 1},{\bf
1})+({\bf 1},{\bf 2},{\bf 2})\,,\quad
T_4:\,\,({\bf 4}, {\bf 1}, {\bf 2}) +2({\bf 6},{\bf 1},{\bf
1})+2({\bf 1},{\bf 2},{\bf 2}) \,,\nn\\
T_3: && 2({\bf 4},{\bf 1}, {\bf 1})+2(\bar{\bf 4},{\bf 1},{\bf
1})+6({\bf 1,1,2})\,,
\label{z6}
\eea
where we have suppressed all the Abelian charges. This model
contains three chiral PS families, two from the $T_1$ sector and
one from the untwisted and $T_{2,4}$ twisted sectors.  Note the
$T_{2,4}$ sectors also contain a $({\bf 4,1,2}) + ({\bf\bar
4,1,2})$ pair which can be used to spontaneously break PS to the
standard model (SM). The complete matter spectrum can be found in
Ref.~\cite{KRZ}. It is natural to identify the two lightest
families with the $T_1$ sector states $({\bf 4,2,1}) + ({\bf\bar
4,1,2})$ located on the $SO(10)$ brane (see Fig.~\ref{fig:z6}).
(In fact, we do not yet understand the dynamics which breaks the
apparent symmetry between these two states.)  The third family is
then identified with the bulk states in $U_1$ and $T_2$. However,
for this identification to be consistent with limits on proton
decay we need $R^{-1} \equiv M_c \gtrsim 10^{16}$ GeV. We return
to this point below when we discuss gauge coupling unification.

Gauge symmetry breaking and matter fields of this model can be
understood in the language of orbifold GUTs. The intermediate
$\mz_3$ model has a GUT group $SO(10)\times SU(2)$ in the
observable sector\footnote{Note, the non-zero roots of the
$SO(10)$ gauge sector are described by momenta ${\bf P} =
(0,0,0,\pm 1,\pm 1, 0, 0, 0) $ (plus all permutations of $\pm 1$
in the last five components).  These satisfy ${\bf P}^2 = 2$ and
${\bf P} \cdot {\bf V_6} = {\bf P} \cdot {\bf W_3} = 0$.  The
weights for the $\bf 16$ and $\bf 10$ dimensional representations
of $SO(10)$ are given by  $(n_1 + \frac{1}{2},n_2 + \frac{1}{2},
n_3 + \frac{1}{2}, \pm \frac{1}{2}, \pm \frac{1}{2}, \pm
\frac{1}{2}, \pm \frac{1}{2}, \pm \frac{1}{2})$ (with an even
number of minus signs for the last five components and $\sum_{i =
1}^3 n_i = 2 \mz$) and  $(n_1, n_2, n_3, \pm 1, 0, 0, 0, 0)$ (plus
all permutations over the last five components with $\sum_{i =
1}^3 n_i = 2 \mz + 1$), respectively. The Wilson line ${\bf W_2}$
preserves $SO(4) \times SO(6)$ where the roots of $SO(4)$ and
$SO(6)$ reside in (4th, 5th) and (6th, 7th, 8th) components of $\bf
P$, respectively.  In addition, $\bf W_2$ distinguishes the Higgs
doublets and triplets.} (modulo Abelian factors), and contains the
following untwisted and twisted-sector matter states in 6d
hypermultiplets
\bea
U\,{\rm sectors}: && ({\bf 16,1})+({\bf 1,2}),\nn\\
T\,{\rm sectors}: && 3({\bf 16, 1})+6({\bf 10,1})+15({\bf
1,2}).\label{matter}
\eea
These matter states are bulk states in the language of orbifold GUTs
(see Fig.~\ref{fig:z3}).  Note that, with the above 6d gauge
sector and matter hypermultiplets, the irreducible 6d $SO(10)$
anomalies cancel \cite{6danomalies}.

The ${\mathbb Z}_2$ orbifold twist ${\bf v}_2$ (represented in the gauge degrees of freedom with the shift $\bf
V_2$) along with the Wilson line $\bf W_2$ generate the two orbifold parities, $P$ and $P'$, in field theory. As
discussed earlier the orbifold parities can be computed for the states in Eq.~(\ref{matter}) using
Eq.~(\ref{PP'}), and they are listed in Table \ref{table1}.
\begin{table}\caption{Parities of the bulk states, {\it i.e.} the states in the gauge, untwisted and $T_2/T_4$
twisted sectors (separated by the horizontal lines respectively).
The sub-indices $\pm$ are intrinsic parities. The multiplicities
represent the number of fixed points in the $G_2$ torus.  All the
states have been decomposed into PS representations.
\label{table1}}
\begin{ruledtabular}
\begin{tabular}{|c|c|cc|c|cc|}
Multiplicities & States & $P$ & $P'$ & States & $P$ & $P'$\\
 \hline
1 & $V({\bf 15,1,1})$ &  $+$ & $+$ &$\Sigma({\bf 15,1,1})$  & $-$ & $-$\\
1 & $V({\bf 1,3,1})$  &  $+$ & $+$ &$\Sigma({\bf 1,3,1})$   & $-$ & $-$\\
1 & $V({\bf 1,1,3})$   & $+$ & $+$ &$\Sigma({\bf 1,1,3})$   & $-$ & $-$\\
1 & $V({\bf 6,2,2})$   & $+$ & $-$ &$\Sigma({\bf 6,2,2})$   & $-$ & $+$\\
\hline
1 & $H({\bf 4,2,1})$   & $+$ & $+$ &$H^c({\bf\bar4,2,1})$   & $-$ & $-$\\
1 & $H({\bf\bar4,1,2})$& $+$ & $-$ &$H^c({\bf 4,1,2})$      & $-$ & $+$\\
\hline
2& $H({\bf 4,2,1})_+$   & $+$ & $-$ &$H^c({\bf\bar4,2,1})_+$   & $-$ & $+$\\
2 & $H({\bf\bar4,1,2})_+$& $+$ & $+$ &$H^c({\bf 4,1,2})_+$      & $-$ & $-$\\
2 & $H({\bf 6,1,1})_+$   & $-$ & $+$ &$H^c({\bf 6,1,1})_+$      & $+$ & $-$\\
2 & $H({\bf 1,2,2})_+$   & $-$ & $-$ &$H^c({\bf 1,2,2})_+$      & $+$ & $+$\\
2 & $H({\bf 6,1,1})_+$   & $-$ & $-$ &$H^c({\bf 6,1,1})_+$      & $+$ & $+$\\
2 & $H({\bf 1,2,2})_+$   & $-$ & $+$ &$H^c({\bf 1,2,2})_+$      & $+$ & $-$\\
1 & $H({\bf 4,2,1})_-$   & $-$ & $+$ &$H^c({\bf\bar4,2,1})_-$   & $+$ & $-$\\
1 & $H({\bf\bar4,1,2})_-$& $-$ & $-$ &$H^c({\bf 4,1,2})_-$      & $+$ & $+$\\
1 & $H({\bf 6,1,1})_-$   & $+$ & $+$ &$H^c({\bf 6,1,1})_-$      & $-$ & $-$\\
1 & $H({\bf 1,2,2})_-$   & $+$ & $-$ &$H^c({\bf 1,2,2})_-$      & $-$ & $+$\\
1 & $H({\bf 6,1,1})_-$   & $+$ & $-$ &$H^c({\bf 6,1,1})_-$      & $-$ & $+$\\
1 & $H({\bf 1,2,2})_-$   & $+$ & $+$ &$H^c({\bf 1,2,2})_-$      & $-$ & $-$\\
\end{tabular}
\end{ruledtabular}
\end{table}

The first embedding removes massless states with orbifold parities $P=-$. Just like in the field theory example,
the $SO(10)\times SU(2)$ gauge group is unbroken. The remaining matter states are $U_1:\,({\bf 16,1})$,
$U_2:\,({\bf 1,2})$, $T_{2}:\,2({\bf 16,1})_++2({\bf 10,1})_-+2({\bf 1,2})_++4({\bf 1,2})_-$,
$T_{4}:\,({\bf\overline{16},1})_-+4({\bf 10,1})_++8({\bf 1,2})_++({\bf 1,2})_- $,
where the sub-indices represent intrinsic parities. The second
embedding, on the other hand, removes states with parities $P'=-$.
It breaks the observable-sector gauge group to the PS group (this
is also identical to the orbifold GUT model).  Finally, massless
matter fields in the untwisted and $T_2,\,T_4$ twisted sectors of
our $\mz_6$ model (Eq.~(\ref{z6})) are the intersections of those
of the two inequivalent embeddings of the ${\mathbb Z}_2$ orbifold
twist, {\it i.e.} the surviving massless states in the 4d effective theory
have orbifold parities $P=P'=+$ which agrees with field
theoretical results.\footnote{It should be noted that the patterns
of gauge symmetry breaking in our models are slightly more general
than those considered in the orbifold GUT literature. Both the $P$
and $P'$ orbifold parities can be realized non-trivially to break
parts of the bulk GUT gauge symmetries. (In the $SO(10)$ orbifold
GUT model \cite{OGUT} and the model presented here, the $P$
parities are trivially realized, in the sense they commute with
all the bulk gauge generators.) In fact, we find additional
three-generation PS models where the intermediate bulk gauge group
is $E_6$, and the two orbifold parities break it to the $SO(10)$
and $SU(6)\times SU(2)$ subgroups at the two fixed points of the
$SO(4)$ lattice. The 4d matter spectra of these models have
similar features to that of the three-generation model presented
here. We relegate the details to Ref.~\cite{KRZ}.}

In the $\mz_6$ model there are also states from the $T_1$ and
$T_3$ twisted sectors. They are localized on the two sets of
inequivalent fixed points of the $SO(4)$ lattice at $0$ and
$\pi R$, and can be properly identified with the brane states
in the orbifold GUT models. From the $SO(4)$ lattice point of
view, these states divide into two sub-sectors, according to their
winding numbers, $n_2=0$ and $n_2=1$, along the direction where
the ${\bf W}_2$ Wilson line is imposed. The set of states with
$n_2=0$ ($n_2=1$) furnish complete representations of the
$SO(10)\times SU(2)$ (PS) group. They are the $SO(10)$ (PS) brane
states in the language of orbifold GUTs (see Figs.~\ref{fig:z2}
and \ref{fig:z6}).

The $T_{1,3}$ twisted-sector states, {\it i.e}, the brane states,
however, are more tightly constrained than their orbifold GUT
counterparts. In orbifold GUT models the only consistency
requirement is chiral anomaly cancelation, thus one can add
arbitrary numbers of matter fields in vector-like representations
on the branes. String models, on the other hand, have to satisfy
more stringent modular invariance conditions \cite{Dixon,vafa}
(which, of course, guarantee the model is anomaly free, up to a
possible Abelian anomaly \cite{LSW}).  These conditions usually
constrain the additional allowed matter in vector-like
representations.   For example, we obtain states transforming in
$({\bf 4,1,1}) + ({\bf\bar 4,1,1})$, $({\bf 1,1,2})$ and $({\bf
1,2,1})$ representations of PS on the PS brane.  We also obtain
states transforming under the hidden gauge group $SO(10)^\prime
\times SU(2)^\prime$.  In addition, the modular invariance
conditions for the gauge twists and Wilson lines also imply that
we cannot project away all the color triplet Higgs $({\bf 6,1,1})$
in our three-generation string model. This feature is different
from that of $SO(10)$ orbifold GUT models. These color triplets do
not necessarily pose the usual doublet-triplet problem as in
conventional 4d GUT models, since in our case the triplets $({\bf
6,1,1})$ and doublets $({\bf 1,2,2})$ have different quantum
numbers (namely, their Abelian charges). Rather than a nuisance,
the color triplets may actually facilitate the breaking of the PS
symmetry to that of the SM. A detailed analysis of the Yukawa
couplings, both at renormalizable and non-renormalizable levels,
and breaking of the PS symmetry will be given in Ref.~\cite{KRZ}.

{\bf 6. Gauge coupling unification}. Finally we determine various
mass scales in our model by requiring gauge coupling unification.
It is highly non-trivial to compute gauge threshold corrections in
string theory \cite{kaplunovsky} in the presence of discrete
Wilson lines, and they are only known numerically in certain cases
\cite{Nilles}. However, in the orbifold GUT limit $R\gg R'\sim
\ell_s$, we only need to keep contributions from the massless and
KK modes (in the ${\bf R}$ direction) below the string scale
$M_s$, and the computation can be done by a much simpler field
theoretical method.\footnote{We impose an explicit cutoff at a
scale $M_s$ which we naturally identify with the string scale.  In
a self-consistent string calculation no explicit cutoff is
necessary.  We do not expect the renormalization group evolution
of the differences of gauge couplings to be affected by our field
theoretic treatment.   On the other hand, the absolute value of
the gauge couplings will obtain scheme dependent threshold
corrections at the cutoff scale.  Only in a self-consistent string
calculation can these corrections be trusted. S.R. thanks H.D. Kim
for emphasizing this point.} Following Ref.~\cite{DDG}, we find
\cite{KRZ}
\be
\frac{2\pi}{\alpha_i(\mu)}=\frac{2\pi}{\alpha_s}+b^{\rm
MSSM}_i\ln\frac{M_b}{\mu} +(b^{\rm PS}_{++}+b_{\rm
branes})_i\ln\frac{M_s}{M_b} -\frac{1}{2}(b^{\rm PS}_{++}+b^{\rm
PS}_{--})_i\ln \frac{M_s}{M_c}+b^{SO(10)}\left(\frac{M_s}{M_c}
-1\right)\,,\label{GQW}
\ee
for $i=1,2,3$, where $M_b$ and $M_c$ are the breaking scale of PS
to the SM and the compactification scale respectively, $\alpha_s$
is the $SO(10)$ gauge coupling at the string scale, and
$M_s\equiv2/\sqrt{\alpha'}=\sqrt{\alpha_s/2}\,M_{\rm pl}$ 
with $M_{\rm pl}\simeq 1.2\times 10^{19}$ GeV the
Planck mass \cite{kaplunovsky}. In this
calculation, we have assumed $M_b\ll M_c$ so that the effect of
symmetry breaking to the KK masses can be neglected. We have also
assumed gauge threshold corrections from particle mass
splittings at the breaking scale $M_b$ are negligible. The third
term on the RHS includes the running due to massless modes as well
as those ``would be'' massless states obtaining mass at $M_b$. The
last term is the contribution of all bulk modes and characterizes
the power-law running of gauge couplings in 5d. Finally, the fourth
term on the RHS takes care of over-countings of the contributions
from massless modes with $+ +$ and $- -$ parities.

In Eq.~(\ref{GQW}), $b_i^{\rm MSSM}=(\frac{33}{5},1,-3)$ is the MSSM beta function coefficient (including one
pair of Higgs doublets). Values of other beta function coefficients and the scales $M_b$ and $M_c$ depend on the
field content below the string scale. As an example we assume 4 (2) bulk hypermultiplets in the ${\bf
16}+{\bf\overline{16}}$ (${\bf 10}+{\bf\overline{10}}$) representations (see Eq.~(\ref{matter})) and 4 pairs of
$({\bf 4,1,1})+({\bf\bar 4,1,1})$ on the PS brane (see Eq.~(\ref{z6})) contribute to the running from $M_c$ or
$M_b$ to $M_s$ (all other states are assumed to get mass at $M_s$).   We then have $b_{++}^{\rm PS}+b_{\rm branes}
=\frac{1}{2}(b_{++}^{\rm PS}+b_{--}^{\rm PS})=(\frac{22}{5},2,2)$ and $b^{SO(10)}=4$. From the
point of view of an effective 4d GUT theory we have the following equations
\be
\frac{2\pi}{\alpha_i(\mu)}\simeq\frac{2\pi}{\alpha_G}+b^{\rm MSSM}_i\ln\frac{M_G}{\mu} + 6\,\delta_{i 3}\,, \label{DRW}
\ee
where the last factor represents the threshold corrections at the GUT scale $M_G \simeq 3 \times 10^{16}$ GeV
necessary to fit the low energy data.  Matching Eqs.~(\ref{GQW}) and (\ref{DRW}) at $M_b$, we find $M_b \simeq
e^{-3/2} M_G \simeq 6.7\times 10^{15}$ GeV and $M_c \simeq e^2 M_G \simeq 2.2\times 10^{17}$ GeV. The string
scale and gauge coupling are $M_s\simeq 2.0\times 10^{18}$ GeV, $\alpha_s\simeq 0.06$. (The latter result is
subject to scheme-dependent threshold corrections at $M_s$ and thus must await a true stringy calculation for
confirmation.) We note that it is safe to identify the two $SO(10)$-brane states in the $\bf 16$ representation
(see Fig.~\ref{fig:z6}) as the lightest two generations of matter, since the compactification scale $M_c$ is
large enough to sufficiently suppress dimension-6 operator contributions to proton decay. (We do not yet
understand the contributions from dimension-5 operators due to color triplet exchanges; they depend on the
precise nature of Yukawa couplings and are left for future investigations.) With the above mass scales, we find
the string dilaton coupling $\sim M_c/\alpha_s M_s\sim {\cal O}(1)$, so the string interaction is in between the
perturbative and non-perturbative regimes and might have very interesting physical implications.

{\bf Acknowledgments}. T.~K.\/ was supported in part by the Grant-in-Aid for Scientific Research (\#14540256) and
the Grant-in-Aid for the 21st Century COE ``The Center for Diversity and Universality in Physics'' from Ministry
of Education, Science, Sports and Culture of Japan. S.~R. and R.-J.~Z. were supported in part by DOE grants
DOE-ER-01545-856 and DE-FG02-95ER40893 respectively. They also wish to express their gratitude to the Institute
for Advanced Study, where initial stages of this work were performed, for partial financial support. Finally,
S.~R. acknowledges stimulating discussions with E. Witten and W. Buchm\"{u}ller and R.-J.~Z. thanks G.~Kane for
encouragement and advice.


\end{document}